# An analysis of stellar interferometers as wavefront sensors

**François Hénault**

CRAL - Observatoire de Lyon, 9 Avenue Charles André, 69561 Saint Genis Laval, France

Abstract: This paper presents the basic principle and theoretical relationships of an original method allowing to retrieve the Wavefront Errors (WFE) of a ground or space-borne telescope when combining its main pupil with a second, decentered reference optical arm. The measurement accuracy of such a "telescope-interferometer" is then estimated by means of various numerical simulations, demonstrating a high performance excepted on limited areas near the telescope pupil rim. In particular, it allows direct phase evaluation (thus avoiding the use of first or second-order derivatives), which is of special interest for the co-phasing of segmented mirrors in future giant telescopes projects. We finally define the useful practical domain of the method, which seems to be better suited for periodical diagnostics of space or ground based telescopes, or to real-time scientific observations in some very specific cases (e.g. the central star in extrasolar planets searching instruments).

## 1. Introduction

The purpose of this paper is to present an unusual measurement method allowing the retrieval of wavefront errors emerging from a ground or space-borne telescope. Such Wavefront Sensors (WFS) are widely spread in the field of astronomy, where many different principles have already been proposed. Among them however, only three or four major concepts have been deeply studied and practically realized. They are the Shack-Hartmann sensor[1-2] based on micro-lens arrays placed at the exit pupil of the telescope (evaluating the first-order derivatives of the wavefront, then reconstructing it digitally), the shearing interferometer[3] also measuring first-order derivatives, the curvature sensor[4-5] estimating second-order derivatives, and the most recently proposed pyramidal WFS[6]. However these methods are hardly applicable for evaluating the piston errors of large segmented mirrors or multi-apertures telescopes, since they are all based on slopes or curvature measurements. In the last decade, this type of mirrors have already been implemented – or are under development – in a few outstanding ground or spaceborne facilities such as the Keck telescopes, the Gran Telescopio Canarias or the James Webb Space Telescope (JWST). In the future they will probably become the building blocks for all Extremely Large Telescope (ELT) projects, and so new alignment and control technologies may be developed jointly.

It must be highlighted that direct phase measurements – including piston errors – are also feasible by means of image-based sensing methods such as phase retrieval or phase diversity procedures[7-8]. For example, they were successfully applied to the Multiple-Mirror Telescope[9] and are the current baseline for the co-phasing of the JWST hexagonal mirrors, using a Gerchberg-Saxton algorithm[10]. However, such techniques usually require significant post-processing time and are not well suited to adaptive optics operation. The "ideal" wavefront sensor should indeed combine the advantages of both methods, i.e. the ability to perform absolute phase measurements in quasi real-time – typically corresponding to 100 Hz time frequency for ground telescopes affected by atmospheric disturbances.

Such techniques have already been proposed by several authors: let us mention the wavefront sensors described by Angel[11-12] and Codona[13], based on a Mach-Zehnder interferometer that could be integrated to extrasolar planets searching instruments. Alternatively, Labeyrie suggested an adaptive holographic correction method[14]. Herein is proposed a quite different principle, inspired from the Michelson stellar interferometer that is famous in the field of astronomy since the beginning of the 20th Century[15]. In the section 2 are described the basic principle and theoretical relationships of the proposed method according to Fourier optics theory. It essentially consists in adding a second, decentered optical arm to the main telescope pupil, thus generating a spatially modulated Point Spread Function (PSF) that is recorded on a detector array. Then the Optical Transfer Function (OTF) of the system can be digitally computed by means of an inverse Fourier transform, providing quantitative information about the telescope WFE (it may be noticed that Takeda[16] followed a comparable approach for interferometric fringe-pattern analysis). The basic principle is validated by several numerical simulations presented in the chapter 3, whereas three possible implementations on real telescopes are described and discussed in the section 4. Finally, the chapter 5 gives a short conclusion about the future potential developments of the method.

## 2. Theoretical principle

The general principle of the proposed method is illustrated in the Figure 1, where the studied telescope is characterized by its exit pupil diameter $D = 2R$ and focal length $F$ – both allowing to define its aperture number $F/D$ – and affected with an unknown wavefront error $\Delta(x,y)$. Then a second optical arm is added into the pupil plane, from where the light emitted by the same sky object is combined to the principal beam. The associated aperture presents the three following characteristics:

- Property n°1: its radius r is sensibly smaller than the telescope radius R (i.e. by one order of magnitude at least)
- Property n°2: it is decentered with respect to the telescope optical axis of a significant quantity denoted B (standing for "baseline". It will be established later that B must be greater than $3R + r$)
- Property n°3: its output beam is converging at the telescope focus and is assumed to be free of phase or alignment defects. In other words it can be considered as diffraction-limited and generating a spherical reference wavefront.

The coherent addition of the complex amplitudes propagated from both apertures will form an interference pattern in the focal plane of the telescope, where the intensities are measured by means of quadratic detectors as in classical stellar interferometers. The measured intensity distribution can also be considered as the PSF of the global system, from which its OTF can be digitally computed through an inverse Fourier transform algorithm. In the following pages, the telescope pupil will be named "main pupil", while the decentered optical arm is the "reference pupil". Thereby we shall seek out analytic expressions of the PSF and OTF of the system, assumed to contain quantitative information about the phase errors in the main aperture.

## 2.1 PSF recorded on the detector

Let us consider the coordinates systems represented on the Figure 1, where:
- Z is the optical axis of the considered telescope

- OXY is the exit pupil plane of the telescope, centered on the main aperture
- O'X'Y' is the focal plane of the telescope, where the PSF intensities will be measured. The point O' coincides with the telescope focus

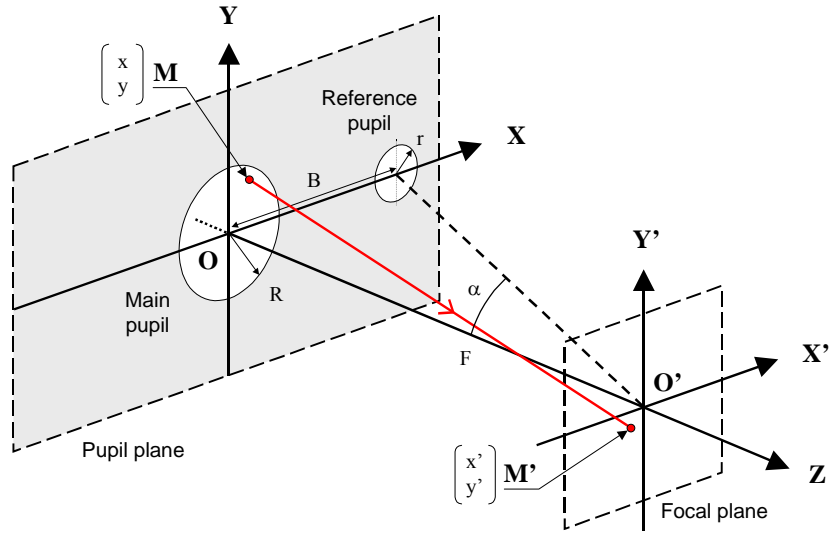

**Figure 1: Coordinates systems**

Let us denote $B_R(x,y)$ the two-dimensional amplitude transmission function of the circular pupil, uniformly equal to 1 inside a circle of radius R, and zero outside of this area – this is the "pillbox" or "top-hat" function. If $\lambda$ is the wavelength of the incoming light (assumed to be monochromatic), the wave emerging from both telescope apertures can be written in the OXY plane:

$$A_P(x, y) = A_R B_R(x, y) \exp[i2\pi \Delta(x, y)/\lambda] + A_r B_r(x - B, y) \quad (1)$$

where $\Delta(x,y)$ is the telescope WFE to be estimated, and $A_R$ and $A_r$ respectively are constant amplitudes in the main and reference pupils (in the most frequent case their numerical values are nearly equal). It can be noticed that no phase error is associated to the reference optical arm according to the aforementioned property n°3. In the frame of scalar diffraction theory and Fraunhofer approximation, the wave generated in the focal plane O'X'Y' is obtained by Fourier transformation of $A_P(x,y)$:

$$A_P'(x',y') = FT[A_P(x,y)] = \iint_{x,y} A_P(x,y) \exp[-i\,2\pi(ux+vy)]\,dx\,dy \qquad (2)$$

with $u = x'/\lambda F$, $v = y'/\lambda F$, and $F$ is the telescope focal length as mentioned above. Then $A_P'(x',y')$ can be linearly developed as follows:

$$A_P'(x',y') = A_R\,FT\{B_R(x,y)\exp[i2\pi\,\Delta(x,y)/\lambda]\} + A_r\,FT[B_r(x-B,y)] \qquad (3)$$

and the Fourier transform of the wave propagated from the main pupil is a complex function that may be rewritten as:

$$FT\{B_R(x,y)\exp[i2\pi\,\Delta(x,y)/\lambda]\} = \pi R^2\,M(u,v)\exp[i\,\Psi(u,v)] \qquad (4)$$

Here $M(u,v)$ and $\Psi(u,v)$ respectively represent the dimensionless modulus and the phase of the incident wave, and the area $\pi R^2$ of the main aperture appears as a normalization factor. The diffracted amplitude from the reference pupil arm can be expressed in a similar way:

$$FT[B_r(x-B,y)] = \pi r^2\,m(u,v)\exp[-i\,2\pi uB] \qquad (5)$$

where $m(u,v)$ is the modulus of the reference wave, equal to the well-known Airy function $2J_1(x)/x$ – see ref. 17, section 8.5 – with no additional phase term since the wavefront is assumed to be free of defects. Hence when combining relations (3) to (5) the complex amplitude $A_P'(x',y')$ formed in the focal plane becomes:

$$A_P'(x', y') = A' \{ M(u, v) \exp[i\, \Psi(u, v)] + C_A C\, m(u, v) \exp[-i\, 2\pi uB)] \} \tag{6}$$

where:
$$A' = A_R \pi R^2 \tag{7}$$

and the constants $C_A$ and $C$ define the contrast ratio between the main and reference pupils:

$$C_A = A_r / A_R \tag{8.a}$$

$$C = r^2 / R^2 \tag{8.b}$$

The intensity distribution $I_P'(x', y')$ in the focal plane is by definition the PSF of the two-aperture optical system, which is equal to the square modulus of $A_P'(x', y')$, i.e. after multiplying by its complex conjugate:

$$I_P'(x', y') = A'^2 \{ M^2(u, v) + C_A^2 C^2 m^2(u, v) \\ + C_A C\, M(u, v)\, m(u, v) [\exp(i\, \Psi(u, v) + i\, 2\pi uB) + \exp(-i\, \Psi(u, v) - i\, 2\pi uB)] \} \tag{9}$$

that can also be written as:

$$I_P'(x', y') = A'^2 \{ M^2(u, v) + C_A^2 C^2 m^2(u, v) + 2 C_A C\, M(u, v)\, m(u, v) \cos[\Psi(u, v) + 2\pi uB] \} \tag{10}$$

It must be noticed that this last expression includes a cosine term proportional to the product $C_A C$, which is indeed very similar to the usual formulation of classical stellar interferometry. Here the modulated component generates a weak intensity fringe pattern added to the telescope PSF. This effect is clearly observable on the Figure 2 where is reproduced an intensity distribution calculated for the first reference WFE studied in the section 3 – see case n°1, "pure defocus".

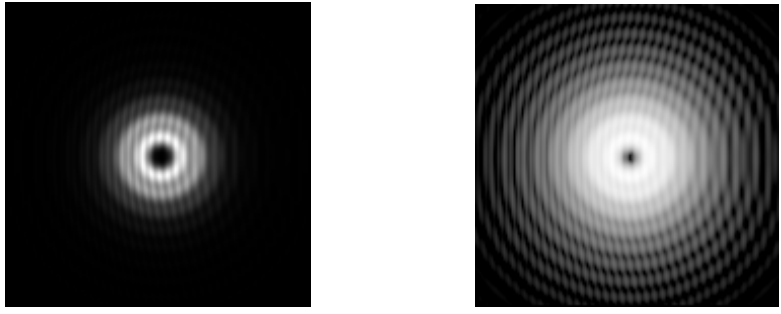

**Figure 2: Example of a telescope PSF modulated by the reference pupil arm (left: linear scale, right: logarithmic scale)**

Having determined the theoretical equations (9-10) of the PSF effectively acquired on the focal plane of the telescope, there only remains pure data-processing tasks. Firstly the OTF must be deduced from the PSF, then a phase retrieval procedure shall be defined.

## 2.2 OTF computation and phase retrieval

Classically the Optical Transfer Function (OTF) of an optical system is equal to the inverse Fourier transform of its Point Spread Function (PSF). However the OTF is generally expressed in terms of the pupil spatial frequencies $f_x$ and $f_y$ where $f_x = x/\lambda F$ and $f_y = y/\lambda F$, and is limited by its cut-off frequency $f_c = 2R/\lambda F$ above which it is uniformly equal to zero. Here the OTF is re-scaled to the pupil plane coordinates x and y by simply ignoring the initial variable substitutions in the focal plane (i.e. $u = x'/\lambda F$ and $v = y'/\lambda F$). Hence from equation (9) the resulting function $C_P(x,y)$ can be split into four major terms:

$$C_P(x,y) = FT^{-1}[I_P'(x',y')] = A'^2 \left\{ C_{P1}(x,y) + C_A^2 C^2 C_{P2}(x,y) + C_A C\, C_{P3}(x,y) + C_A C\, C_{P4}(x,y) \right\} \quad (11)$$

Although an analytical development of the $C_{P1}(x,y)$ to $C_{P4}(x,y)$ terms is straightforward in Fourier optics theory, it is fully detailed into the Appendix in order to define the exact normalizing factors and the resulting contrast ratio. Therefore it can be shown that:

$$C_P(x,y) = A'^2 \left\{ \frac{1}{\pi R^2} O_R(x,y) + C_A^2 C^2 \frac{1}{\pi r^2} O_r(x,y) + C_A C \frac{1}{\pi R^2} \frac{1}{\pi r^2} \left\{ \left( B_R(x,y) \exp\left[ i \frac{2\pi}{\lambda} \Delta(x,y) \right] \right) \otimes B_r(x+B,y) \right\} \right. \\ \left. + C_A C \frac{1}{\pi R^2} \frac{1}{\pi r^2} \left\{ \left( B_R(-x,-y) \exp\left[ -i \frac{2\pi}{\lambda} \Delta(-x,-y) \right] \right) \otimes B_r(x-B,y) \right\} \right\} \quad (12)$$

where $O_R(x,y)$ and $O_r(x,y)$ respectively are the OTFs of the main and reference pupils, each being considered individually, and the symbol $\otimes$ denotes a convolution product. Even if they are scaled to pupil coordinates instead of spatial frequencies, these functions are similar to the typical "Chinese-hat" profiles described in ref. 17, section 9.5. Hence when normalizing $C_P(x,y)$ such that its maximal value (at $x = y = 0$) is equal to 1, we finally obtain:

$$C_P(x,y) = \frac{1}{1 + C_A^2 C} \left\{ O_R(x,y) + C_A^2 C\, O_r(x,y) + C_A C \left( B_R(x,y) \exp\left[ i \frac{2\pi}{\lambda} \Delta(x,y) \right] \right) \otimes \frac{B_r(x+B,y)}{\pi r^2} \right. \\ \left. + C_A C \left( B_R(-x,-y) \exp\left[ -i \frac{2\pi}{\lambda} \Delta(-x,-y) \right] \right) \otimes \frac{B_r(x-B,y)}{\pi r^2} \right\} \quad (13)$$

In the Figure 3 is provided a schematic representation of the modulus of the complex function $C_P(x,y)$, which is indeed the Modulation Transfer Function (MTF) of the two-aperture optical system (see also the Figure 4 corresponding to the defocus example in section 3). It shows that both the MTF and OTF are constituted of four terms being interpreted as follows:

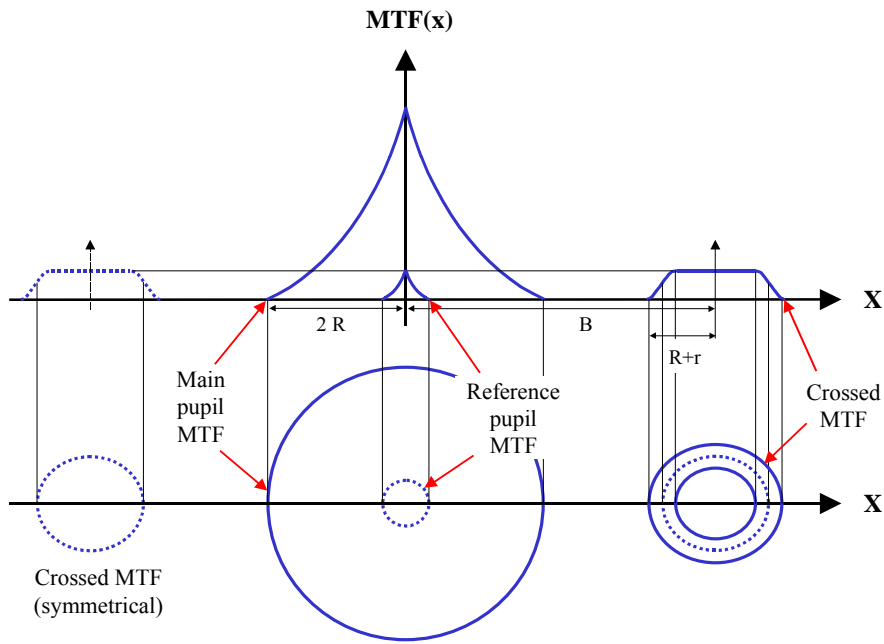

**Figure 3: MTF of the global system (schematic)**

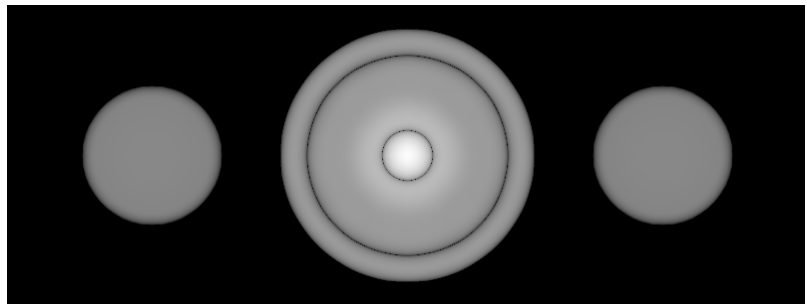

**Figure 4: Typical example of a global MTF (logarithmic scale)**

1) The first term proportional to $O_R(x,y)$ is nothing else than the OTF of the main aperture, i.e. the transfer function of the telescope taking into account its limited diameter and transmitted wavefront error $\Delta(x,y)$

2) Likewise, the second term is proportional to $O_r(x,y)$ that is the OTF of the reference pupil. Given that it is proportional to the contrast ratio C and that its radius 2r is significantly smaller than R (see the property n°1), it can be assumed that this function has negligible influence with respect to the first term

3) The third component is by far the most interesting: firstly because it is a crossed term between both main and reference pupils appearing as a convolution product and involving a complex function whose phase is directly related to the searched wavefront error $\Delta(x,y)$. Secondly, because it is shifted of an algebraic distance -B along the X-axis, meaning that it can easily be separated from the other terms provided that the effective value of B is high enough to prevent any spatial overlap between them. As the maximal – or "cut-off" – radius of $O_R(x,y)$ is 2R and the diameter of the convolution product is at most equal to 2(R+r), we deduce that the condition to be strictly fulfilled is:

$$B > 3R + r \tag{14}$$

4) Finally the fourth term appears as a rigorous replication of the previous one, rotated of 180 degrees around the optical axis and centered on the +B coordinate along the X-axis, thus being symmetrical with respect to point O. In theory this crossed-symmetrical term does not contain any additional information.

In order to isolate the crossed term between the main and reference pupils in equation (13), we shall multiply it by the pillbox function $B_{R+r}(x+B,y)$ of radius R+r and centered on the -B coordinate, then shift the whole result of the opposite quantity +B toward the origin O. This leads to the following relation:

$$\frac{1+C_A^2 C}{C_A C} B_{R+r}(x,y) C_P(x-B,y) = \left( B_R(x,y) \exp\left[ i\frac{2\pi}{\lambda} \Delta(x,y) \right] \right) \otimes \frac{B_r(x,y)}{\pi r^2} \tag{15}$$

This last result proves that the third component of the OTF is equal to the convolution product of the complex amplitude transmitted by the main telescope aperture with a circular pillbox function $B_r(x,y)$ uniformly equal to 1 inside the radius r. Obviously the equation (15) could be resolved by means of a deconvolution process. However such algorithms often require extensive computing times and are not easily applicable to quasi real time operation. Another solution consists in referring to the first property stipulating that r is negligible with respect to the main telescope radius R. Thus the function $B_r(x,y)/\pi r^2$ can be approximated to the Dirac distribution $\delta(x,y)$, and the convolution product in Eq. (15) might be neglected – it must be noticed that this approximation is all the more valid as the r/R ratio tends to decrease. Consequently it may be assumed that:

$$B_R(x,y)\exp\left[i\frac{2\pi}{\lambda}\Delta(x,y)\right] \approx \frac{1+C_A^2 C}{C_A C} B_{R+r}(x,y)\, C_P(x-B,y) \qquad (16)$$

Neglecting the real multiplying coefficients on both sides of equation (16), the WFE retrieval formula can finally be expressed as:

$$\Delta(x,y) \approx \frac{\lambda}{2\pi}\operatorname{Arctan}\left\{\frac{\operatorname{Im}[B_{R+r}(x,y)\,C_P(x-B,y)]}{\operatorname{Real}[B_{R+r}(x,y)\,C_P(x-B,y)]}\right\} \quad \operatorname{mod}[\lambda] \qquad (17)$$

where Im[ ] and Real[ ] respectively stand for the imaginary and real parts of a complex number. This last relationship concludes the theoretical study, since we have then defined a procedure allowing phase or WFE reconstruction from the PSF generated at the focal plane of a two-aperture telescope system. However, it must be emphasized that the absolute accuracy of the method shall suffer from an intrinsic measurement error originating from the "delta approximation" of Eq. (16). Therefore its actual performance needs to be quantified from a few practical examples: this is the main scope of the numerical simulations presented in the following section.

## 3. Numerical simulations

In order to validate the global WFE retrieval procedure and to estimate its intrinsic measurement error, we developed a simulation routine following the main steps detailed below:

1) Firstly, the wavefront error $\Delta(x,y)$ to be measured is imported from an external file. It is considered as a reference, to be finally compared with the retrieved data in step n°9. Four different reference WFEs are studied here as described hereafter. Their spatial sampling is typically around 100 x 100, as indicated in the Table 1

2) Phase errors proportional to $\Delta(x,y)$ are added to the main aperture of the telescope, which is included together with the reference pupil into a complex array $A_P(x,y)$ according to the relation (1). The reference aperture is typically de-centered of 400 pixels from the origin and sampled by 10 x 10 pixels. Phase errors can also be associated to it, although this was not undertaken here. The dimension of the computing array is adjustable, but experience showed that the best results are obtained from a 1024 x 1024 sampling. Larger array sizes only result in prolonged computing time without any observable gain on the measured performance

3) The complex amplitude distribution $A_P'(x',y')$ in the telescope focal plane is then evaluated via a direct Fourier transformation of the 1024 x 1024 complex array following the relation (2)

4) $A_P'(x',y')$ is multiplied by its complex conjugate in order to simulate the PSF of the global system measured by the CCD detector matrix, that is $I_P'(x',y')$. No photon or read-out noises are added. At this stage the optical simulation is completed and the following steps consist in pure data-processing

5) According to the general relation (11), the OTF of the system is computed in the pupil plane by means of an inverse Fourier transform

6) The resulting complex distribution $C_P(x,y)$ is multiplied by a pillbox function of radius R+r centered on the -B coordinate, and the result is shifted of +B along the X-axis according to the relation (15). Thus the crossed OTF term is extracted from $C_P(x,y)$

7) The estimated wavefront error of the telescope is derived from the previous data according to the relation (17)

8) As equation (17) can only provide numerical values ranging from 0 to $\lambda$ (or from $-\lambda/2$ to $+\lambda/2$), a phase unwrapping algorithm must be applied to the estimated WFE if these limits are exceeded. Such procedures have been extensively studied and optimized in the frame of laser-interferometer metrology and are widely spread in scientific literature. Here we utilized the algorithm originally proposed by Takeda[16] adding some of the improvements suggested by Roddier[18]

9) Finally, the recovered and eventually unwrapped WFE is compared with its reference (see step n°1) by direct subtraction of their two-dimensional maps.

Numerical simulations were conducted for a telescope having an exit pupil diameter 2R = 500 mm and a focal length F = 5 m, thus having an effective aperture number of 10. The baseline B between the main and reference apertures is always equal to 1 m with $A_r = A_R$, meaning that $C_A = 1$ according to the relation (8.a). All the other input parameters are provided in the Table 1 as well as the most significant numerical results of the simulations. Before discussing the numbers compiled in this Table, the four different considered wavefront errors are described below. Each of them is illustrated by an image strip where the grey-levels are linearly scaled to the obtained Peak-to-Valley (PTV) figures.

**Table 1: Synthesis of simulation results**

|  |  | **Case n°1**: Pure defocus | **Case n°2**: Low spatial frequency defects | **Case n°3**: Segmented mirrors defects (piston) | **Case n°4**: Random turbulence defects |
|---|---|---|---|---|---|
| **INPUT PARAMETERS** | Pupil sampling | 129 x 129 | 99 x 99 | 129 x 129 | 99 x 99 |
|  | Wavelength ($\mu m$) | 0.6328 | 0.6328 | 1 | 0.6328 |
|  | Reference pupil diameter (mm) | 50 | 50 | 25 | 40 |
| **REFERENCE WAVE-FRONT ERROR** | PTV ($\lambda$) | 0.983 | 1.727 | 0.997 | 3.395 |
|  | RMS ($\lambda$) | 0.283 | 0.333 | 0.324 | 0.925 |
| **RETRIEVED WAVE-FRONT ERROR** | PTV ($\lambda$) | 0.898 | 1.620 | 0.997 | 3.281 |
|  | RMS ($\lambda$) | 0.273 | 0.326 | 0.323 | 0.920 |
| **DIFFERENCE MAP** | PTV ($\lambda$) | 0.090 | 0.123 | 0.182 | 0.251 |
|  | RMS ($\lambda$) | 0.018 | 0.013 | 0.011 | 0.016 |
| **ERROR RATIO (%)** | PTV | 9.1 | 7.1 | 18.2 | 7.4 |
|  | RMS | 6.4 | 4.0 | 3.5 | 1.7 |

<u>Case n°1</u> (Figure 5): This is a "pure defocus" defect, meaning that the tested telescope does not suffer from any manufacturing error or phase disturbance, but that the focal plane – or the CCD detector itself – is shifted by a small amount along the Z optical axis. Therefore the WFE is reduced to a spherical term (here equivalent to one wavelength PTV, see Table 1). In the Figure 5 are displayed from left to right the original wavefront error in the main telescope pupil, the modulus of the crossed OTF term obtained at the step n°6 of the numerical model, the reconstructed wavefront error in step n°7, and the two-dimensional difference map between the original and estimated WFEs. As also illustrated by the cross-sections of Figure 6, it can be noticed that the major discrepancies are located near the contour of the main aperture, where a sudden slope discontinuity occurs. The width of the corrugated zone corresponds to the radius r of the reference pupil. This inherent measurement error directly originates from the approximation of Eq. (16), because the convolution product tends to integrate useful information located inside the telescope aperture radius R with meaningless data found outside of it.

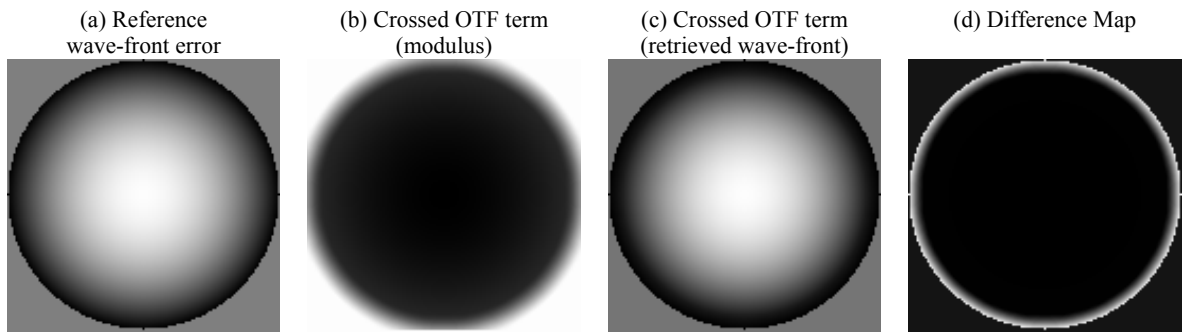

**Figure 5: Reference (a) and reconstructed WFEs (c) of a pure defocus defect, and their two-dimensional difference-map (d). Grey-levels are scaled to PTV values indicated in Table 1**

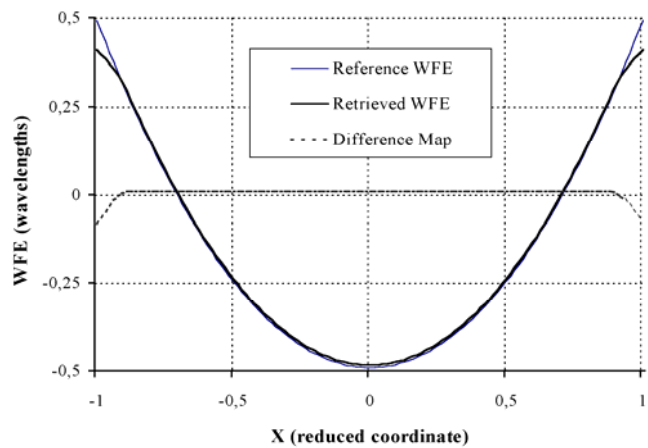

**Figure 6: Cross-sections of the reference, reconstructed and difference WFEs for a pure defocus defect**

Case n°2 (Figure 7): These are low spatial frequency defects, engendered either by polishing errors and mechanical deformations of the telescope mirrors, or by an incorrect alignment between them: typical examples are coma and astigmatism aberrations. Here the WFE amplitude is higher than one wavelength, thus the unwrapping algorithm mentioned in the step

n°8 had to be used. As in the previous example, the difference map (d) reveals the typical slope discontinuity inherent to the convolution product.

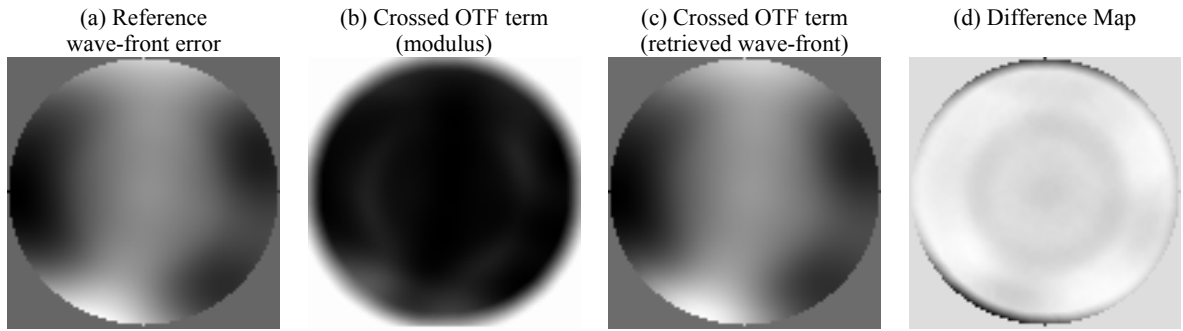

**Figure 7: Reference (a) and reconstructed WFEs (c) of low spatial frequency defects, and their two-dimensional difference-map (d). Grey-levels are scaled to PTV values indicated in Table 1**

Case n°3 (Figure 8): Here we consider the particular case of segmented mirrors that will equip the future ELTs. In the Figure 8 is shown a main pupil constituted of seven hexagonal facets, some of which are affected with piston errors presenting the following characteristics: the first and central segment is assumed to be the reference optical surface having a zero piston (i.e. the surrounding mirror petals are aligned with respect to it). Then, turning clockwise from the Y-axis, the other reflective segments present successive piston errors of $\lambda/2$, $\lambda/4$, $-\lambda/4$, $-\lambda/2$, $-\lambda/4$ and $\lambda/4$ (a). The radius r of the reference pupil had to be reduced with respect to both previous examples (see Table 1) in order to enhance the spatial resolution near the edges of the mirrors. It can be seen that the modulus of the crossed OTF term (b) reproduces a blurred picture of the transmission map in the main pupil, while the piston errors are actually retrieved on the whole mirror segments (c), excepting restricted areas located near the boundaries of each petal (d). Although the case was not studied here, it may be assumed that a slightly enlarged reference pupil would still allow to restore the piston errors.

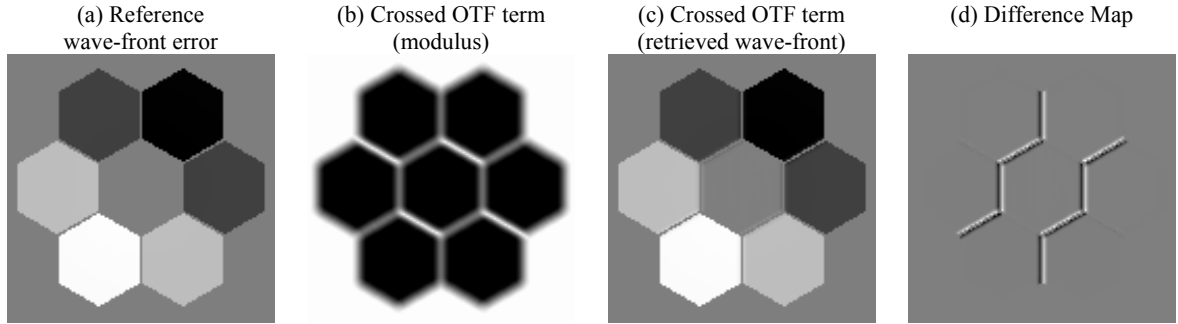

(a) Reference wave-front error  (b) Crossed OTF term (modulus)  (c) Crossed OTF term (retrieved wave-front)  (d) Difference Map

**Figure 8: Reference (a) and reconstructed WFEs (c) of segmented mirrors defects, and their two-dimensional difference-map (d). Grey-levels are scaled to PTV values indicated in Table 1**

<u>Case n°4</u> (Figure 9): These are random defects, representative of optical path disturbances encountered during the propagation into turbulent atmospheric layers and observed in the exit pupil plane of ground telescopes (i.e. the astronomical seeing). Such effects are nowadays corrected by advanced adaptive optics systems requiring fast frequency WFS[1-6]. They often are high-amplitude defects (equivalent to several wavelengths), thus the use of phase unwrapping algorithms is absolutely necessary. Nevertheless, as the other digital procedures are well suited to quasi real-time processing (inverse Fourier transform, filtering and phase extraction), we have studied if the proposed method can be employed for this purpose. Here again the radius r of the reference pupil had to be reduced, but in lesser proportions than for the prior example: the actual reason was to avoid large phase variations averaged on the reference aperture, inducing MTF values close to zero – as seen on the upper region of Figure 9-(b) – and subsequent phase extraction errors occurring at step n°7 of the numerical model. The difference map (d) also shows the typical slope discontinuity already noticed in the first and second examples.

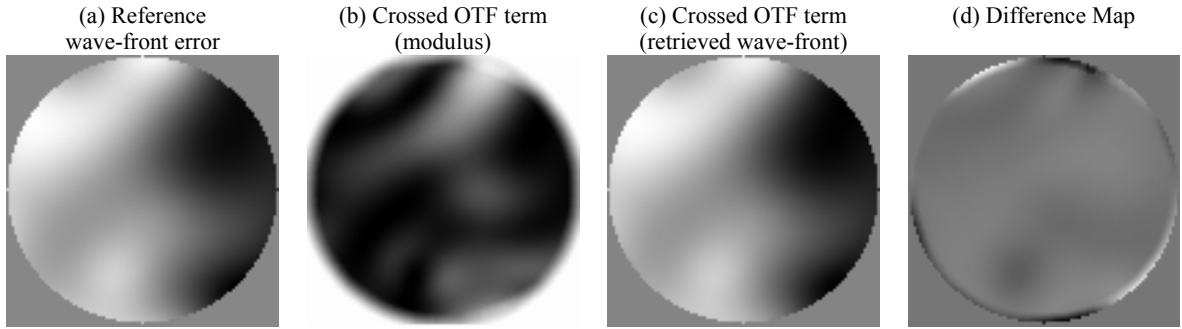

**Figure 9: Reference (a) and reconstructed WFEs (c) of random defects, and their two-dimensional difference-map (d). Grey-levels are scaled to PTV values indicated in Table 1**

All the numerical results obtained for these four typical cases are compiled in the Table 1, where are indicated the PTV and RMS values of the reference and reconstructed WFEs, and of their difference maps as well as global estimates of the error percentage. The obtained accuracy is ranging from around $\lambda/11$ to $\lambda/4$ PTV and from $\lambda/88$ to $\lambda/55$ RMS, and thus always stays within the diffraction limit of the main telescope. It may seem at a first glance that the proposed method is less accurate than other current WFE measurement techniques such as Shack-Hartmann or image-based sensing. However it must be pointed out that the major discrepancies originate from the slope discontinuity generated by the convolution product near the main pupil rim. For cases n° 1, 2 and 4, it can be assumed that a much better accuracy is achieved over a circular sub-aperture of radius R-r excluding the corrugated areas. However such a "pupil reduction" process is only appropriate to some particular cases (for example the defocus or low spatial frequency defects that can realistically be extrapolated by Zernike polynomials), and this is the reason why it was not studied here. In other words such improvements of the measurement accuracy are feasible, but can only be envisaged on a case-by-case basis.

But perhaps one of the most decisive advantages of the technique was confirmed by the third example: it shows that direct information about the searched wavefront is accessible, instead of utilizing its slopes or curvatures. This allows a simultaneous retrieval of the co-phasing errors of segmented telescope mirrors, which is of prime interest in view of future ELT projects. Therefore the method can be thought as an efficient tool for regular controls of the telescope alignment and performance – in other words, an active optics or image-based sensing technique. Furthermore, its capacities in the field of adaptive optics remain to be assessed, despite of the encouraging results presented in the last example.

Although this study should be pursued in order to determine the optimal values of the system parameters (and particularly those of the reference pupil diameter), we will now provide a few examples of practical implementations on an existing telescope. This is the purpose of the following section.

## 4. Practical implementation and other considerations

Having asserted the general principle of an original wavefront sensing method and defined its basic theoretical relationships (section 2), then demonstrated its capacities through various numerical simulations (section 3), we describe hereafter three potential examples of implementation on a real telescope.

The first configuration is illustrated on the Figure 10 and shows a typical telescope from the Cassegrain family where both primary mirror (M1) and secondary mirror (M2) have conic surfaces. The telescope is considered as the main pupil of the two-aperture optical system, while the second reference arm is schematically represented at the bottom of Figure 10: it is composed of two off-axis mirrors having the same curvatures and conic constants than the primary and secondary mirrors, being de-centered of a quantity equivalent to B along the X-axis (i.e. they are direct extensions of the M1 and M2 optical surfaces). Hence this configuration can be seen as a Fizeau stellar interferometer where one of the two apertures has been significantly reduced, and folding mirrors are added to the reference arm for compactness purpose. The latter also comprises a Delay Line (DL) maintaining an equal Optical Path Difference (OPD) between both telescope pupils, for example driven by a Piezoelectric Transducer (PZT) tube – a necessary condition to observe the modulated PSFs shown in Figure 2. It may be noticed that due to the different number of reflective surfaces encountered along each optical arm, the transmitted amplitudes $A_R$ and $A_r$ will not be equal, thus justifying a posteriori the introduction of the contrast ratio $C_A$ in the section 2.

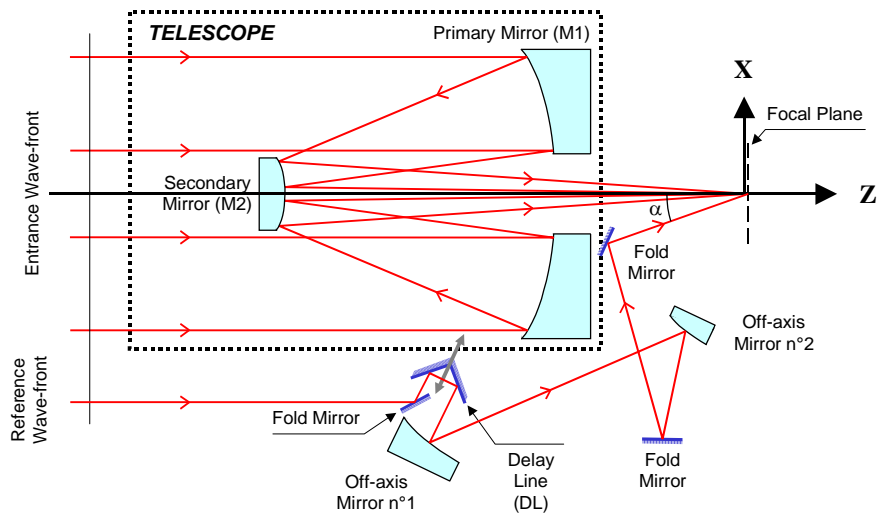

**Figure 10: Implementation on a real telescope. The Fizeau configuration**

The second example slightly deviates from the previous one. As can be seen in the Figure 11, the reference pupil arm is no longer composed of the same type of mirrors than the main telescope, but of a dioptric (or eventually catadioptric) telephoto lens system having the same focal length than the telescope, again incorporating folding mirrors and a PZT delay line. This configuration cannot be related to the Fizeau family since the continuity of the optical surfaces between both interferometer arms vanishes. Conversely, it cannot be seen as a Michelson-like stellar interferometer where the light is usually collected from several telescopes of the same type, then transported towards a beam combiner via relay optics. Therefore it will be called the "intermediate" configuration. It is indeed very similar to the first one, eventually offering the advantages of a more compact optical layout and reduced bulkiness.

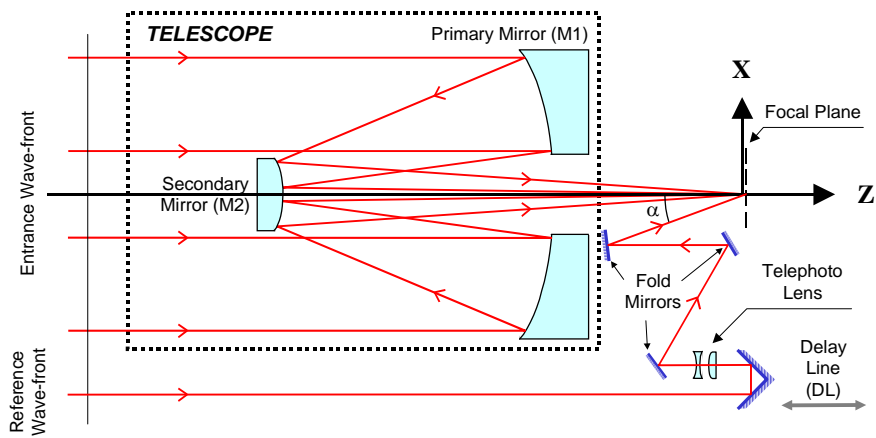

**Figure 11: Implementation on a real telescope. The intermediate configuration**

The third and last proposed arrangement (see Figure 12) radically differs from the previous ones. It is indeed a Michelson configuration involving two different telescopes coupled by fiber optics as in the FLUOR interferometer[19]: the first one still corresponds to the main aperture where the WFE is to be measured, while the second "Auxiliary Telescope" (AT) can be of intermediate size between the main and reference pupil diameters. The light collected by the AT is then focused at the entrance of an optical fiber and finally injected into the main beam from an off-axis angle equivalent to B, via focusing optics and two folding mirrors. Several options can be envisaged in order to equalize the OPDs between both interferometer arms: for example a classical delay line (similar to those mentioned in the preceding configurations) could be implemented near the input or output sides of the fiber, which would certainly require additional optics. But the best solution probably consists in incorporating the DL to the optical fiber itself, as represented in the Figure 12. This can be realized by means of a ceramic cylinder PZT as demonstrated by Reynault[20].

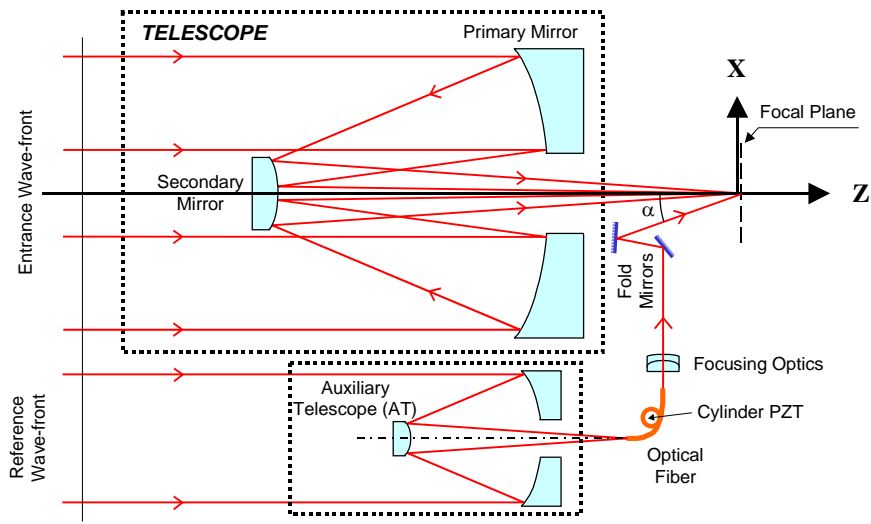

**Figure 12: Implementation on a real telescope. The Michelson configuration**

Two important peculiarities of the "Michelson" configuration must be highlighted below:

- In this case the numerical values of the wave amplitudes $A_R$ and $A_r$ might significantly differ from each other, and the diameter of the auxiliary telescope could be adjusted so that $A_r > A_R$, implying a contrast ratio $C_A$ higher than 1 and thus an amplified PSF modulation term – proportional to $C_A$, see the relation (10). Therefore the measurement accuracy of the wavefront could be improved. It must be noticed however that this may be detrimental to scientific exposures since the diffraction effects originating from the reference pupil are proportional to $C_A^2$.

- A potential difficulty may be to maintain a sufficient image quality over the whole surface of the auxiliary telescope, that is supposed to provide a spherical reference wavefront. For ground facilities it entails that the atmospheric seeing is well corrected, either by an adaptive optics system or by means of passive WFE filtering techniques. The optical fiber itself could be used for that purpose[21].

Among many other possibilities, the three schematic configurations here above described demonstrate that the concept of such "telescope-interferometers" may not be just a dream, even if substantial effort will be required before they can actually be realized. In this perspective, a few basic questions about their domain of practical use are still to be discussed.

<u>Wavelength range</u>: this crucial question concerns the spectral characteristics (peak wavelength and width) of the observed luminous source. At a first glance it may seem that the reconstructed phase do not depend on the considered wavelength $\lambda$ since the OTF is scaled to pupil dimensions by means of the intermediate variables (u,v) – see the relations (4) to (10). Unfortunately this is untrue since the PSFs actually recorded on the detector array depend on the (x',y') coordinates and not on (u,v). Therefore the inverse Fourier transform of equation (11) has to be calculated for a given reference wavelength $\lambda_0$, e.g. the peak wavelength of the observed object. Then at different wavelengths the OTFs will be re-scaled by geometrical similarities of ratio $\lambda_0/\lambda$, consequently scrambling the information available on the main pupil area. Hence the light source must either be a spectral line or filtered on a narrow spectral bandwidth (typically a few percents). The choice of the peak wavelength $\lambda_0$ does not seem to have a critical influence on the accuracy of the method, but rather on its measurement range as discussed below.

<u>Maximal wavefront amplitude</u>: it has already been highlighted in section 3 that if the amplitude of the estimated wavefront is greater than the useful wavelength $\lambda_0$, a phase-unwrapping algorithm must be implemented. But in some cases this may lead to information loss, for example when the co-phasing errors of segmented mirrors must be evaluated (see section 3, example n°3). In that case higher reference wavelengths $\lambda_0$ could be selected by means of spectral filtering. For very-high amplitude phase errors infrared sources might be used if they can provide sufficient energy.

Angular source size and Field of View (FoV): these are two other essential issues. It readily appears that in order to avoid PSF modulation drops the angular size of the observed source should not exceed the resolving power of the equivalent stellar interferometer, that is $\lambda/B$ – where B must comply with the relation (14). Practically, this means that the candidate sky objects may be restricted to natural stars of a few tenths of milli-arcseconds diameter, and that the method is not compatible with laser guide stars. Besides, the useful FoV of the system will be subject to the classical limitations of stellar interferometry, which is a vast theme far beyond the scope of this paper. Let us simply remind that the Fizeau configuration is reputed for having the widest FoV, which may give preference to the two first proposed arrangements ("Fizeau" and "intermediate"). Even in that case however, it seems difficult to process simultaneously different sky objects without any readjustment of the delay line. It can then be concluded that unless the DL system is deeply modified, the method is not suitable for multi-conjugate or ground-layer adaptive optics, and is mostly applicable to unresolved central stars, such as those observed by extrasolar planets searching instruments.

Delay Line range and accuracy: following the previous hypothesis (i.e. negligible FoV requiring a moderate DL compensating range), the measurement uncertainty should not be very sensitive to OPD equalization errors: their major effect is indeed to add a global piston to the estimated WFEs, from which it will finally be subtracted digitally. Therefore only a moderate positioning accuracy should be required for the delay line, and its practical design could be greatly simplified and not very expensive.

Science and reference pupil baseline: the point is here to define the optimal value of the baseline B separating both telescope apertures: it has been shown in section 2 that the inequality (13) has to be strictly respected in order to prevent any overlap between the different OTF terms. Moreover a small margin should naturally be added to this lower limit. However it can be assumed that larger baselines will probably have little beneficial influence on the WFE spatial resolution and measurement accuracy, because the system does not act as a real stellar interferometer. On the contrary, the precision could even be decreased for very long baselines, since the PSF spatial modulation may become hardly observable by means of CCD detector arrays. Therefore it seems preferable to keep the baseline parameter B close to its lower limit (i.e. 3R + r).

Image quality of the reference arm: when referring to Twymann-Green or Fizeau interferometers commonly used for optical surfaces metrology, it might be expected that the reference telescope arm must fulfil stringent image quality requirements (as suggested by the property n°3 in section 2). Sensitivity analyses were carried out using the numerical model described in the section 3, and showed that wavefront errors of $\lambda/4$ PTV (in both tilt and defocus) are equivalent to half the intrinsic measurement accuracy of the method. Therefore the reference telescope should be diffraction-limited according to the Rayleigh criterion, and does not really need to be totally free of phase or alignment defects as initially assumed in section 2. Obviously some more detailed image quality budgets will need to be assessed in the future.

Photometry and noise aspects: flux and noise issues probably represent the most critical areas of the method, at least for an adaptive optics operating mode. Given the aforementioned characteristics of the observed source (i.e. limited to a narrow spectral bandwidth and unresolved by the interferometer baseline), it might be feared that poor Signal-to-Noise Ratios (SNR) are barely achieved on a limited number of sky objects. However this key question requires a dedicated study that will be the scope of another paper, thus only general considerations are given here. Let us consider for example the case of photon noise: referring to Eq. (10) the total collected flux is proportional to $A'^2$, so the photon noise will depend on $A'$. However the useful signal from which the phase will be extracted is equal to $2 A'^2 C_A C$, implying that the effective SNR is proportional to $\pi r^2 C_A = S C_A$ where $S$ and $C_A$ respectively are the collecting area and concentration ratio of the reference telescope. When considering the largest existing facilities or the future ELT projects, the diameter of the reference telescope should attain a few meters, provided that it is assisted by its own adaptive optics system. Thus the SNR naturally increases with the size of the considered facility. In addition, a major enhancement should result from the "Michelson" configuration incorporating larger Auxiliary Telescopes with significantly improved concentrating ratios (i.e. $C_A > 1$).

## 5. Conclusion

In this paper was discussed the theory and expected performance of a wavefront sensing technique inspired from stellar interferometry. Firstly, we described the basic principle and theoretical relationships allowing WFE or phase retrieval on a ground or space-borne telescope when combining its main pupil with a reference off-axis optical arm. The measurement accuracy of such a "telescope-interferometer" was then estimated from various numerical simulations, demonstrating a high performance excepted on limited areas located near the telescope pupil rim. In addition, it was shown that direct phase evaluation is achieved (thus avoiding the use of first or second-order derivatives), which is of prime interest for the co-phasing of segmented mirrors in view of innovative projects such as the future ground-based ELTs or the Terrestrial Planet Finder (TPF). We finally attempted to define the useful practical domain of the method, which seems well suited to periodical diagnostics of ground or space telescopes (in other words, an active optics or image-based sensing technique) or to real-time scientific observation in a few specific cases (e.g. the central star in extrasolar planets searching instruments).

It seems clear however that some important issues mentioned in the previous section must be examined in depth. Using our numerical model, a thorough analysis of the measurement errors will be conducted, including noises and non-linearity of the CCD detector, useful spectral range and angular size of the observed sky objects, scintillation effects, etc. This should be the scope of another paper currently under preparation, and of future tradeoffs with respect to classical WFS or image-based sensing techniques. But it already appears that some experimental results would be of great interest in order to assess the validity and predicted performance of the method. A gradual approach divided into three steps is suggested below:

- A laboratory test-bench should be firstly developed, allowing to check the measurement accuracy and to assess several physical characteristics of the envisaged light sources (such as spectral width, angular size and brightness), thus answering to the major open questions in section 4.
- If proved successful, a real telescope-interferometer could be built by adding a reference optical arm to an existing telescope of moderate size (around 1-2 meter diameter), and tested during on-sky observations.
- In order to evaluate the ultimate performance of the method, the last step would consist in implementing it on a large-scale telescope. For example, experiments carried out on the Large Binocular Telescope (LBT) or on the Very Large Telescope Interferometer (VLTI) would respectively allow to compare the Fizeau and Michelson configurations described in the previous section.

## Appendix: computation of $C_P(x,y)$

According to the relation (11), $C_P(x,y)$ is a linear combination of four different terms that are analytically developed below.

### Expression of $C_{P1}(x,y)$

This term appears as the inverse Fourier transform of $M^2(u,v)$ that is the PSF of the main telescope aperture. It can be rewritten as follows:

$$C_{P1}(x,y) = FT^{-1}[M^2(u,v)] = FT^{-1}[M(u,v)\exp(i\,\Psi(u,v)) \times M(u,v)\exp(-i\,\Psi(u,v))] \qquad (A.1)$$

and then expressed as the convolution product – here denoted $\otimes$ – of two complex conjugated functions:

$$C_{P1}(x,y) = FT^{-1}[M(u,v)\exp(i\,\Psi(u,v))] \otimes FT^{-1}[M(u,v)\exp(-i\,\Psi(u,v))] \qquad (A.2)$$

According to the relation (4) and to the properties of Fourier transformations for complex conjugates, $C_{P1}(x,y)$ is consequently equal to:

$$C_{P1}(x,y) = \frac{1}{\pi R^2}\left(B_R(x,y)\exp\left[i\frac{2\pi}{\lambda}\Delta(x,y)\right]\right) \otimes \frac{1}{\pi R^2}\left(B_R(-x,-y)\exp\left[-i\frac{2\pi}{\lambda}\Delta(-x,-y)\right]\right) \qquad (A.3)$$

It can be noticed that the expression (A.3) is proportional to the auto-correlation function $O_R(x,y)$ of the complex amplitude in the main pupil, which is another classical definition of the OTF[17]. Knowing that $O_R(x,y)$ is always normalized such that $O_R(0,0) = 1$, we find that:

$$\left(B_R(x,y)\exp\left[i\frac{2\pi}{\lambda}\Delta(x,y)\right]\right)\otimes\left(B_R(-x,-y)\exp\left[-i\frac{2\pi}{\lambda}\Delta(-x,-y)\right]\right)=\pi R^2\,O_R(x,y) \tag{A.4}$$

and then, combining both previous relationships:

$$C_{P1}(x,y)=\frac{1}{\pi R^2}O_R(x,y) \tag{A.5}$$

## Expression of $C_{P2}(x,y)$

The same demonstration than for the previous term easily leads to:

$$C_{P2}(x,y)=\frac{1}{\pi r^2}O_r(x,y) \tag{A.6}$$

## Expression of $C_{P3}(x,y)$

According to the relation (11) the expression of $C_{P3}(x,y)$ is:

$$C_{P3}(x,y)=FT^{-1}[M(u,v)\,m(u,v)\exp(i\,\Psi(u,v)+i\,2\pi uB)] \tag{A.7}$$

that can be decomposed into a multiple convolution product:

$$C_{P3}(x,y)=FT^{-1}[M(u,v)\exp(i\,\Psi(u,v))]\otimes FT^{-1}[m(u,v)]\otimes FT^{-1}[\exp(i\,2\pi uB)] \tag{A.8}$$

Then, from the relations (4-5) and the basic properties of Fourier transforms:

$$C_{P3}(x,y)=\frac{1}{\pi R^2}\left(B_R(x,y)\exp\left[i\frac{2\pi}{\lambda}\Delta(x,y)\right]\right)\otimes\frac{1}{\pi r^2}B_r(x,y)\otimes\delta(x+B) \tag{A.9}$$

where $\delta(x,y)$ is the Dirac distribution, and $C_{P3}(x,y)$ may finally be written as:

$$C_{P3}(x,y) = \frac{1}{\pi R^2} \frac{1}{\pi r^2} \left\{ \left( B_R(x,y) \exp\left[ i \frac{2\pi}{\lambda} \Delta(x,y) \right] \right) \otimes B_r(x+B, y) \right\} \quad (A.10)$$

*Expression of $C_{P4}(x,y)$*

Here again, the same demonstration than for $C_{P3}(x,y)$ and the Fourier transform properties of complex conjugates readily lead to:

$$C_{P4}(x,y) = \frac{1}{\pi R^2} \frac{1}{\pi r^2} \left\{ \left( B_R(-x,-y) \exp\left[ -i \frac{2\pi}{\lambda} \Delta(-x,-y) \right] \right) \otimes B_r(x-B, y) \right\} \quad (A.11)$$